# Reciprocity or community?

## Different cultural pathways to cooperation and welfare

Anna Gunnthorsdottir

*University of Arizona and University of Iceland*

Palmar Thorsteinsson

*Central Bank of Iceland\**

**Corresponding author:**
Anna Gunnthorsdottir
annagu@hi.is
Social Sciences, room 213
1145 E South Campus Dr
Tucson, AZ 85721

**Acknowledgments:** This work was supported by Australian Research Council (https://www.arc.gov.au/) Discovery grant DP1095395
An earlier version was presented in the Center for the Philosophy of Freedom seminar series at the University of Arizona on April 15, 2021. The authors thank Lee Cronk, Saura Masconale, Mary Rigdon, David Schmidtz, and Gylfi Zoega for helpful comments.

Keywords: Cooperation, culture, pro-sociality, cultural differences, experiment

JEL codes**:** C91, C92, Z10, Z18

\*The views expressed are those of the author and do not necessarily reflect those of the Central Bank of Iceland**.**

# ABSTRACT

In a laboratory experiment we compare voluntary cooperation in Iceland and the US (N=156 in a convenience sample of university students). We furthermore compare the associated thought processes across cultures. The two countries have similar economic performance, but survey measures show that they differ culturally. Our hypotheses are based on two such measures, Inglehart's cultural world map and Knack & Keefer's scale of civic attitudes toward large-scale societal functioning. We prime the participants with different social foci, emphasizing in one a narrow grouping and in the other a larger social unit. In each country we implement this using two different feedback treatments. Under group feedback, participants only know the contributions by the four members of their directly cooperating group. Under session feedback they are informed of the contributions within their group as well as by everyone else in the session. Under group feedback, cooperation levels do not differ between the two cultures. However, under session feedback cooperation levels increase in Iceland and decline in the US. Even when contribution levels are the same members of the two cultures differ in their motives to cooperate: Icelanders tend to cooperate unconditionally and US subjects conditionally. Our findings indicate that different cultures can achieve similar economic and societal performance through different cultural norms and suggest that cooperation should be encouraged through culturally tailored suasion tactics. We also find that some decision factors such as Inequity Aversion do not differ across the two countries, which raises the question whether they are human universals.

# 1. INTRODUCTION

## 1.1 CULTURAL ECONOMICS

Culture is "*a set of customary beliefs and values that ethnic, religious, and social groups transmit fairly unchanged from generation to generation*" (Guiso, Sapienza, & Zingales, 2006, p. 23). Constrained by the human genetic makeup and cultural group selection, culture gives rise to diversity in values and behavior among humans, across geography and history (Cronk, 1999; Soltis, Boyd & Richerson, 1995). Passed on within the family or community, it shapes a specific set of internalized norms and preferences, lenses through which the world is interpreted, and thoughts and feelings about what one experiences or observes (Storr & John, 2019; Henrich & Ensminger, 2014; Tabellini, 2010; Fernández, 2008; Cronk 1999; Cavalli-Sforza & Feldman, 1981).

Since culture influences how individuals process interpersonal exchange it impacts social capital (Guiso, Sapienza & Zingales 2004; Fukuyama, 2001; Coleman, 1988) and is thus an indirect determinant of economic and societal performance(Gershman, 2017).[1] For example, norms of generalized trust and trustworthiness reduce or eliminate transaction costs (Coleman), and many transactions are outright impossible without broad norms of honesty and reciprocity (Tabellini, 2010; Putnam, Leonardi & Nanetti, 1993; Banfield, 1958). Similarly, strong norms of voluntary cooperation make a social unit cohesive and competitive (Torgler, 2004; Ostrom, 1990).

Here, we experimentally examine cultural differences in voluntary cooperation. We explore not only behavior but also the underlying motives. We compare Iceland and the United States, two established Western democracies with developed market economies and similar levels of welfare (OECD, 2020; World Bank, 2019). These commonalities should remove confounds associated with the economic and political environment such as familiarity with a market economy or wealth, both of which have been tied to expressions of culture (Henrich & Ensminger, 2014; Inglehart & Welzel, 2005). Superficial parallels notwithstanding, the two countries differ along all established survey measures of culture known to us. We prime participants with different social foci, emphasizing either one's directly cooperating group or the wider social unit. We discover cultural differences in how the primes impact cooperation, and in the underlying motives to cooperate. Our findings suggest that due to motivational differences, welfare enhancing pro-sociality should be encouraged differently across these different cultures.

---

[1] Seminal classics on the link between history, culture, social capital, and welfare include Banfield's (1958) case study of a South Italian village and Putnam et al.'s (1993) comparison of Southern and Northern Italy. Since then, the link between culture and welfare has been established in greater detail (e.g., Knack and Keefer,1997; La Porta et al., 1997; Tabellini, 2010; Algan & Cahuc, 2010; see Gershman, 2017, and Spolaore, 2014, for overviews).

## 1.2 MEASURING CULTURE WITH SURVEYS AND EXPERIMENTS

In addition to qualitative (ethnographic) approaches, culture is frequently assessed via extensive cross-national surveys of values and norms such as the World Values Survey (**WVS**) or the European Values Survey (**EVS**). Readers might be familiar with survey-based quantifications of culture such as Inglehart's cultural map (Inglehart & Baker, 2000) or Hofstede's (2001) cultural dimensions. Economists often know Knack & Keefer's (1997) index of national civic attitudes. Other established scales include GLOBE (House et al., 2004), Schwartz values (Schwartz, 1992), and Group-Grid (Douglas & Wildavsky, 1983). Recently, simple, widely used experimental economics paradigms have been proposed as complements or even alternatives to cross-cultural surveys, especially as instruments for the cross-cultural assessment of narrowly delineated aspects of culture (Thöni, 2019; Gershman, 2017; Henrich & Ensminger, 2014; Guiso, Sapienza & Zingales, 2006).

In economics experiments, all participants anonymously play the same, easy to grasp, context-free game, and have comparable monetary incentives.[2] This way, the experimental environment is stripped of the legal and reputational pressures that often constrain behavior and constrain it differently across different cultures. Cross-cultural experiments thus greatly reduce social desirability effects (ref) and help reveal culture-specific internalized norms. Unlike the broad culture surveys, economics experiments target defined aspects of culture that impact social capital, such as perceptions of fairness, norms of reciprocity, or standards around cooperation and the voluntary provision of public goods.

Since culture is a shared mindset, cultural differences translate into behavioral differences only indirectly (Cronk, 1999, p. XI). It is conceivable that in some settings, different cultures behave similarly. This might sound like an argument against cross-cultural experiments which, after all, assess behavior. However, experimental cultural comparison is not limited to outward behavior. It is possible to dig deeper and econometrically analyze decision factors and thus, motives as well as values deeply or even unconsciously held. Econometric comparison of decision processes across cultures is to the best of our knowledge a new approach. It is the approach we take in this paper. Our focus is on motives underlying voluntary cooperation, an aspect of culture closely linked to social capital. Our econometric approach is inspired by Ashley, Ball & Eckel's (2010) analysis of cooperative motives among US students. We next discuss the role of cooperation around public goods for a social unit's success. This is followed by a description of how experiments assess voluntary cooperation.

---

[2] Within-culture experiments suggest that as long as the payoff functions are linear transformations of each other, stake sizes are of minor importance (Kocher, Martinsson & Visser, 2006). However, the added complexity of cross-cultural experiments requires special thoroughness. To avoid possible confounds, the stakes in the game should be equivalent across locations. See Section 2.6 for our approach to this.

## 1.3 VOLUNTARY PUBLIC GOODS PROVISION, COOPERATION, AND WELFARE

Public goods produced via voluntary cooperation are widespread and diverse.[3] They include the quality of the environment, water supply, shared pastures or fishing grounds (Hardin, 1968), tax revenue (Camerer, 2003, p. 46), public safety and order, or defense. A rationally self-interested actor does not voluntarily contribute to a public good but "snatches a selfish benefit" (Samuelson, 1954, p. 389) from others' contributions. A culture's ability to instill norms that curb such selfishness is a significant determinant of its effectiveness, the welfare of its members, and even its long-term survival (Ahn & Ostrom, 2008; Boyd, & Richerson, 1985, 2002; Knack & Keefer, 1997, pp. 27-29; Soltis, Boyd & Richerson, 1995). A large body of field observations (e.g. Ostrom, 1990) has generated a rich set of hypotheses about which cultural aspects help or hinder voluntary cooperation. Laboratory experiments with the Voluntary Contribution Mechanism (VCM) allow testing these hypotheses under controlled conditions.

## 1.4 THE VOLUNTARY CONTRIBUTION MECHANISM (VCM)

The Voluntary Contribution Mechanism's (**VCM**; Isaac, McCue & Plott 1985) simplicity and versatility have made it the work horse to examine voluntary cooperation experimentally.[4] We describe the most common version of the game, used in this paper.

There are $i = 1, ..., n$ symmetric group members. Each player $i$ has a monetary endowment $e$. Each $i$ contributes $x_i \in [0, e]$ to a group account which represents the public good and leaves the remainder $(e-x_i)$ in a personal account. The return on investment in the group account differs from the personal account's return. For simplicity and without loss of generality, set the return from the personal account to one. The group account contributions by all $n$ group members are summed up and multiplied by a factor $g$ which represents the benefits from cooperation, before being equally divided among all $n$ members. $g/n$ is the Marginal Per Capita Return (**MPCR**) to each group member from an investment in the group account. Each group member's payoff is:

---

[3] For seminal contributions see Hardin, 1968; Olson 1965; Samuelson, 1954; Pigou 1932/2017, p. 182).
[4] A Google Scholar search for "Voluntary Contribution Mechanism" on August 25, 2021, returned 2770 matches.

$$\Pi_i = (e - x_i) * 1 + \left(\sum_{i=1}^{n} x_i\right) * g/n \tag{1}$$

If $1 < g < n$ aggregate welfare is maximized if all participants contribute their entire endowment $e$ to the group account. However, a rationally self-interested player's dominant strategy is to contribute nothing while still getting an equal share of the total in the group account. Parameterized this way, the VCM models the tension between individual and collective interest.[5]

**A typical VCM experiment**

Groups of size $n = 3$ to $n=5$ are most common (Zelmer, 2003). The number of groups, **G,** most often varies between two and five. To eliminate reputation concerns the game is played anonymously: It is common knowledge that no player knows the others' identity, and never will. Participants are visually separated from each other. To allow them to learn about the game and each other's choices the game is repeated at least ten times, often much more. In each round, after all players have allocated their funds each player $i$ receives private feedback with her earnings for the round, and information about the contributions of the other group members.[6] In most VCM experiments, participants are randomly assigned to a new group after each round, effectively resulting in a series of one-shot interactions. "Partners" treatments where players stay in the same group in all rounds are less common.[7] After the last round each participant is privately paid her cash earnings.

## 1.5 STYLIZED VCM RESULTS FROM WESTERN UNIVERSITY STUDENTS

The overwhelming majority of VCM experiments has been conducted with university students in Western industrialized countries. These studies have produced three robust results (Zelmer, 2003; Ledyard 1995, Davis & Holt, 1993).

1. Cooperation levels exceed the dominant strategy equilibrium of non-contribution by all. In the first round, the mean contribution is about ½$e$ (Camerer, 2003, p. 46).

---

[5] If $g > n$ it is both individually and collectively optimal to invest everything in the group account. With $g < 1$ it is both individually and socially optimal that everyone keeps their endowments in their private accounts.

[6] Small variations in the nature of the feedback message (for example, information about either aggregate or disaggregated contributions in one's group) have a negligible impact on contributions especially when subjects are assigned to a new group at each round (Cox & Stoddard, 2015).

[7] Comparisons between these "partners" treatments and the more common "strangers" treatments where groups are randomly composed at each round have yielded unclear, even contradictory results (for an overview, see Andreoni & Croson, 2008).

2. Over repeated rounds, mean contributions gradually approach the equilibrium. The decay is mainly because those who initially contributed get discouraged by free riders in their midst (Gunnthorsdottir, Houser & McCabe, 2007; Page, Putterman & Unel, 2005).
3. There is substantial variability in the contributions, especially in the early rounds (Gunnthorsdottir et al.; Kurzban & Houser, 2005; Isaac, Walker, & Thomas, 1984).

### 1.6 CROSS-CULTURAL VCM EXPERIMENTS

Recently, experiments with participants from outside the common subject pools have indicated that the seemingly robust findings from VCM experiments reflect not human universals but rather, a unique culture: Across history and geography, Western university students are outliers in lifestyle and values. In an extensive discussion, Henrich, Heine & Norenzayan (2010) label these subjects as **WEIRD** (Western, Educated, Industrialized, Rich, Democratic). Outside Western universities and even between WEIRD subject pools, substantial cultural differences have since been discovered for example in the perception of what is fair, willingness to punish transgressors, and in the levels of voluntary cooperation (Herrmann, Thöni & Gächter, 2008; Henrich et al. 2001; 2005).

Cross-cultural economics experiments have often been exploratory, with the choice of subjects determined by researchers' access to them. They often simply aimed to check whether cultural differences existed, without prior directional hypotheses. There are exceptions, mostly for the Ultimatum Game,[8] and to a lesser extent also for the VCM. The most prominent deductive approach to cultural differences in experimental games is arguably the Market Integration Hypothesis which suggests that a society's economic practices shape its values and norms (see Ensminger & Henrich, 2014, for an overview; see also Chen & Tang, 2009; Henrich et al. 2005, p. 811; Gurven, 2004). We next provide a brief overview of cross-cultural VCM studies whose results can be summarized as follows:

1. Cross-cultural variation in contributions is substantial.
2. Two of the robust findings from VCM experiments with WEIRD subjects generalize: There are large individual differences in contributions especially in the early rounds, and contributions decline over repeated rounds.

---

[8] The Ultimatum Game assesses fairness perceptions: Player 1 proposes how to divide an amount of money, but Player 2 can refuse the proposed division. If Player 2 refuses both players get nothing.

**1.6.1 INTERNATIONAL COMPARISONS**

A pioneering examination of cultural diversity in VCM contributions is Henrich et al.'s (2001; 2005) study of six geographically diverse small-scale societies with traditional economies, from self-sufficient agriculturists to hunter-gatherers with a high level of cooperation. In their one-round experiments[9] the mean contributions range from 22% among self-sufficient family farmers to 65% in close-knit tribal societies with a tradition of voluntary cooperation such as pooling food. Among Blackwell & McKee's (2010) student subjects Russians contribute the most followed by Kazachs, with the US a distant last. Gächter & Herrmann (2009) report nation-level effects in the VCM contributions of Swiss and Russian university students: Contributions did not differ between each country's universities, but Russian students cooperated less than Swiss students. Gächter, Herrmann & Thöni (2010) compare the cooperation rates of 1120 university students from 16 countries including some that are not democratic or are developing. Their study demonstrates the impact of culture on cooperation rates: Contributions are similar between related cultures but differ between locations that are culturally distinct. (Section 1.7.1 describes their study in more detail.) Ehmke, Lusk & Tyner (2010) compare university students in France, two US states, China and Niger. Contribution patterns in the two US locations and in China are near identical, and French contributions are somewhat similar to them. Oddly, Nigerien cooperation levels start below those of the other countries but do not decline over rounds.

**1.6.2 COMPARISONS BETWEEN "WEIRD" REGIONS**

Our study falls into this category. Both Iceland and the US are established democracies. Their per capita GDP (PPP adjusted) has long been close (OECD, 2020; World Bank, 2019). Yet, as our study will also show, noteworthy cultural differences can exist between WEIRD cultures. Ockenfels & Weimann's (1999) report that East German students contribute much less than West German students, presumably because their post-WWII ideological and economic divergence impacted civic attitudes, a process that is often slow to reverse (Gershman, 2017; Alesina & Fuchs-Schündeln, 2007). The East German mean first round contribution (roughly 22%) is about as low as the contribution by self-sufficient Amazonian farmers reported by Henrich et al. (2005). In Bigoni et al.'s (2016) experiments, Southern Italians cooperate less than Northern Italians; this is not surprising since the two regions have long differed historically, economically, and culturally (see e.g., Putnam et al., 1993). Castro (2008) finds that British students contribute more than Italian students. Weimann (1994) reports lower cooperation among US students than German students in a "partners" VCM. In a study that is not cross-cultural but whose econometric approach inspires the current study, Ashley et al. (2010) conducted an econometric analysis of the structure of contribution decisions in two classic VCM experiments with US students in two different states.[10] The coefficients of the variables subjects consider in their contribution decision are similar in both locations. However, Eckel, Harwell & Castillo (2015) do find a state-based difference among US subjects: within the state of Texas students contribute similarly and more than their peers in other states.

---

[9] A one-round experiment is equivalent to the first round of a multi-round experiment where subjects get re-grouped in each round since in both settings, contributing to motivate other group members to do the same is not possible.

### 1.6.3 VCM EXPERIMENTS WHERE NO CULTURAL DIFFERENCES WERE FOUND

Ehmke et al.'s (2010) cross-cultural study mentioned in Section 1.6.1 contains an interesting null result: US contribution levels do not differ from China's even though there is no doubt that these cultures are quite different. In a meta-analysis of VCM experiments conducted in East Asia and the US, Pang & Bowles (2006) similarly report no significant differences. Brandts, Saijo & Schram (2004) find no statistically significant country effects between the contributions of Spanish, Dutch, Japanese and US students. Henrich et al. (2010) suggest that such behavioral parallels might be due to university students increasingly sharing a global culture. An alternative explanation is that since culture is a shared world view that impacts behavior but does not equal behavior (Cronk, 1999) culturally different thought processes can result in either different behavior or in superficially similar behavior. Recall for example that the cooperation rates Henrich et al. (2001; 2005) found among self-sufficient tribal South American farmers resemble those of Ockenfels & Weimann's (1999) East German students.

In the next section, we develop hypotheses about cultural differences between Iceland and the US regarding observed cooperation levels as well as the underlying cognitions. In Section 3 we return to the question of whether different cultures can attain similar levels of efficiency through different culturally driven mental processes.

### 1.7 HYPOTHESES BASED ON SURVEY MEASURES OF NATIONAL CULTURE

Cross-cultural economic experiments are relatively new. This paper aims to establish and clarify their role as alternative, targeted instruments to measure culture. We therefore base our hypotheses on two established survey measures, Inglehart's world map of cultures (Inglehart & Baker, 2000), and Knack & Keefer's (1997) index of civic attitudes. These two scales lead us to hypothesize that VCM contribution levels and the underlying motivational processes differ between the US and Iceland (Inglehart & Baker), and that in Iceland, a focus on the wider social unit increases cooperation more than in the US (Knack & Keefer).[11]

---

[10] University of Arizona (Isaac & Walker, 1988)/University of Wisconsin (Andreoni, 1995).

***Figure 1***.  *The 2010-2014 Inglehart-Welzel world map of cultural values (World Values Association, 2020a).*

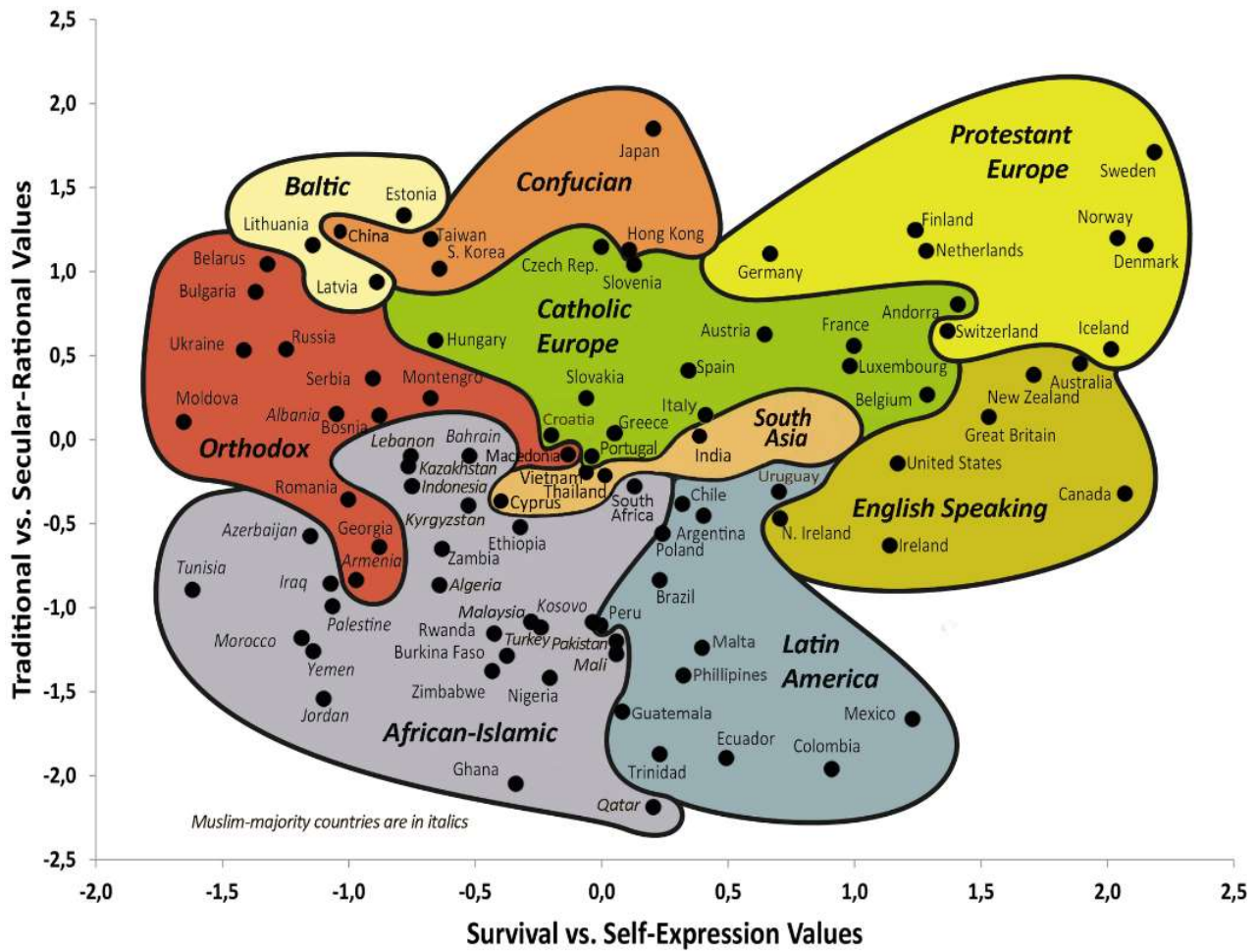

---

[11] We use these two scales because they offer the clearest hypotheses about voluntary cooperation and the associated thought processes. Hofstede's six-dimensional scale for example is harder to link to VCM behavior because both the dimension of Individualism and the dimension of Masculinity might predict behavior in VCM experiments, as pointed out by Blackwell & McKee (2010).

## 1.7.1 THE CULTURAL WORLD MAP

In the world map of values (Fig. 1) introduced by Inglehart & Baker (2000), two orthogonal axes are constructed from responses to the WVS.[12] The vertical axis captures the tension between "traditional" and secular-rational values. A "traditional" culture favors nationalism, religion, and authority, while a secular-rational culture favors agnosticism and science. The horizontal axis captures the tension between survival and self-expression: Security-focused survival values prioritize safety and predictability while post-materialistic emancipative self-expression values favor individualistic unfolding (Welzel, 2013; Inglehart & Welzel, 2005). Most importantly for us here, Fig. 1 shows that national cultures form clusters of related cultures based on a factor analysis of survey responses. The clusters illustrate culture's persistence over time: Religious commonalities and historic links shape present-day cultural similarity. Inglehart and collaborators have extensively written about change in certain values as economies develop. Especially, with increasing wealth cultures tend to move from the traditional, security focused lower left corner of the map toward the secular-rational, individualist upper right corner of the map (Welzel, 2013; Inglehart & Welzel, 2005; Inglehart & Baker, 2000; Inglehart, 1997). Such movement notwithstanding, the clusters have remained consistent.[13] Fig. 1 depicts the 2014 version of the map since our data were collected in late 2014 and in early 2015.

**Findings from VCM experiments**

In a large international VCM study in 16 locations spanning six of Inglehart's culture clusters Gächter, Herrmann & Thöni (2010) report that the clusters account substantially for the observed variation in contributions while contributions are remarkably similar *within* clusters. Their analysis is general and focused on overall sources of variability between and within clusters, *not* on bilateral differences between clusters. We hypothesize, somewhat tentatively:

**Hypothesis 1:** *Since Iceland and the US belong to different culture clusters, contribution levels differ*[14]*between the two countries.*

While culture impacts behavior it does so via values and perceptions (Cronk, 1999). Consider for example the striking classic experiment by Morris, Nisbett & Peng (1995) on how differently Far

---

[12] The WVS questions are subject to regular review and occasional updates (Welzel, 2013, Ch.2). See e.g., World Values Survey (2020a) for the methodology, or Kistler, Thöni, & Welzel (2017) for a more concise description.
[13] See World Values Survey (2020b), https://www.worldvaluessurvey.org/WVSContents.jsp, for a visual representation of changes in the map over the years, and how the culture clusters have remained consistent.
[14] The reader will appreciate the rationale for a difference hypothesis rather than directional hypotheses. Linking survey measures of culture to experiments is a new approach. The sole study this hypothesis is based on is Gächter et al.'s (2010) who do not report bilateral differences between clusters.

Eastern and Western cultures interpret even simple visual stimuli. Theirs and similar[15] studies underscore how different cultures focus on different aspects of a situation and interpret it differently. Also consider the homogeneity in the interpretation of stimuli *within* cultures (Morris et al.). Regarding situations that call for cooperation, in their econometric decomposition of factors that subjects consider when deciding on their VCM contribution, Ashley et al. (2010) report coefficients that are surprisingly similar for US students in two different states. We are aware of only one study that attempts to explore what underlies differences in VCM contributions across cultures:[16] Cherry et al. (2008) report that in the US conditional cooperator types are more common than in Austria or Japan, suggesting that different cultures view the decision to cooperate differently.

**Hypothesis 2:** *Since Iceland and the US belong to different culture clusters subjects in the two countries process their contribution decision through different cognitive lenses.*

### 1.7.2 NORMS OF CIVIC COOPERATION

In a seminal study, Knack & Keefer (1997) account for economic growth with a population's pro-sociality, assessed via five questions from the WVS. They call their explanatory variable CIVIC. Table 1 shows that Iceland is more "CIVIC" than the US, since overall, the anti-social acts listed are less tolerated there. While the CIVIC difference between the two countries seems small, Knack & Keefer report that seemingly small differences in CIVIC have a significant impact on growth (p. 1257).

**Findings from VCM experiments**

Herrmann, Thöni & Gächter (2008) conducted VCM experiments with an added punishment option where each participant could reduce the earnings of any other group member at a cost to self. They report that CIVIC is associated with a willingness to punish free riders. It seems plausible that a culture where free riders get punished by their peers values cooperation and pro-sociality. Punishment might be an attempt to strategically force free riders to cooperate, but it might also be due to a simple moral aversion to behavior deemed anti-social.

A closer look at CIVIC reveals that, while it captures attitudes toward cooperation, the questions are not about a directly cooperating team such as in a VCM, but rather, about pro-social behavior

---

[15] For an overview of similar studies see Nisbett & Masuda, 2003.
[16] While Cherry et al.'s study uses a VCM, their approach differs from ours: They use the Strategy Method (Selten, 1967), then classify individual participants by type (reciprocator, free-rider) based on their responses.

toward broad public goods where cooperation is not directly reciprocated but is based on a moral imperative to behave in a pro-social manner. We therefore put forward the following hypothesis:

**Hypothesis 3:** *Since Iceland is more "CIVIC" than the US, Icelandic subjects contribute more than US subjects when the information they receive during the experiment focuses them on the wider community rather than just on their directly cooperating team.*

*Table 1*. Mean responses to single statements that make up the CIVIC score, and CIVIC scores per country.[17]

| How justified is it to: | Iceland (EVS 2010) | US (WVS 2010-14) |
|---|---|---|
| **Evade taxes if the opportunity arises** | 1.98 (v234) | 1.91 (v201) |
| **Accept public assistance that you have no right to receive** | 1.50 (v233) | 2.30 (v 198) |
| **Dodge fares on public transport** | 2.72 (v 247) | 2.59 (v199) |
| **SUM** | 6.2 | 6.8 |
| **CIVIC** | 23.8 | 23.2 |

WVS/EVS question numbers in parentheses, year of data collection in parentheses.
For individual responses, a response of 1 means "never justified" while 10 means "always justified".
SUM is the summary score of the three questions.
CIVIC is computed as (30 – SUM). CIVIC ranges from 3 to 30; higher numbers reflect higher prosociality.

## 2. METHOD

### 2.1 PARTICIPANTS

The US participants were recruited from the student population at [*a large US state university*]. They were invited for a two-hour experiment with a US$ 5 payment for showing up on time, and further earnings contingent upon the decisions by themselves and others during the experiment.

---
[17] Only three of Knack & Keefer's original five questions are part of recent WVS waves. The questions about the following two activities are no longer included in the WVS: "Keeping money that you have found", and "failing to report damage you have accidentally done to a parked vehicle".

Students who had previously participated in a VCM experiment were excluded. The Icelandic subject pool was the students at *[a large Icelandic university]*. Their show-up fee was ISK 1000. No students were excluded from the invitation since no VCM experiments had been conducted there.

## 2.2 LOCATIONS, FACILITIES, AND SOFTWARE

The US experiments were run in a decision laboratory with networked computers separated by permanent blinders. The Icelandic experiments took place in a multi-purpose computer lab with temporary cubicles installed for the experiment. The software, in Visual Basic, is available from the authors upon request.

## 2.3 EXPERIMENTAL PARAMATERS

For reasons of comparability, we opted for parameters frequently used in VCM experiments (see Zelmer, 2003, for common VCM parameters). The return from the private account was 1. The sum of investments in the group account was multiplied by $g=2$. The group size $n$ was four so that the MPCR was 0.5 (see Eq. 1). There were twelve subjects in each session, divided into $G=3$ groups which were randomly re-composed at each round. The number of rounds was $t=80$. At the start of each round each subject received an endowment of $e=100$ tokens. See Appendix A for the instructions.

*Table 2*. *Experimental design*

|  | Standard ("team") feedback. | Session ("community") feedback |
|---|---|---|
| **Iceland** | 3 sessions with 12 subjects each | 4 sessions with 12 subjects each |
| **US** | 3 sessions with 12 subjects each | 3 sessions with 12 subjects each |

## 2.4 DESIGN

The design (Table 2) is a fully crossed 2x2. The experiments were run in two high-income established democracies that differ culturally. This allows us to separate the impact of socio-economic development (Welzel, 2013; Inglehart & Welzel, 2005) from long standing cultural values. In both locations there were two treatments that differ in the end-of round information a subject received:

- The *group-level("team") feedback* treatment has standard VCM end-of-round information: After each round, a subject was reminded of his own contribution to the individual and group account in that round, the sum of what the other three group members together had contributed, his total earnings in that round, and those earnings broken down according to their source (individual account/group account). Additionally, each participant saw boxes with the group contributions of all *n*=4 members of his group, with his own contribution highlighted. This setup, where participants know the extent of cooperation within their group, is a simple model of a social unit such as a team, organized for direct, mutually beneficial cooperation.
- *Session-level ("community") feedback* is designed to assess the effect of focusing subjects not only on their directly cooperating team but also on a wider social unit (Hypothesis 3). In addition to what subjects saw during team feedback, at the end of each round each subject was shown boxes with the contributions not only within her team but by all session participants. The boxes were grouped into three clusters corresponding to the grouping in that round. A subject's own contribution was again highlighted so that her team, and her own contribution, were clearly recognizable. This treatment models a community comprised of organizations with mobile membership, where all "community members" are aware of each other's actions even though they are not necessarily directly impacted by them.

The instructions (Appendix A), in English in both locations, were identical across feedback treatments except for the second-to-last paragraph where the treatment-dependent end-of-round feedback is explained.

### 2.5 PROCEDURE

Since lectures and textbooks at *[a large Icelandic university]* are often in English the experiments were conducted in English in both locations, according to the same written protocol, by the same experimenters not previously known to the participants. At the start subjects got paid their show-up fee and were assigned to cubicles at random. Next, an experimenter announced that the experiment had started and that no communication, verbal or otherwise, was permitted. Subjects were instructed to place their bags under their seats, shut off their phones, read and sign the consent forms,[18] and to raise their hand if they had questions which an experimenter would address individually and privately.

---

[18] Approved by the [*large US state university*] Institutional Review Board and accepted by the University of Iceland Student Registrar's Office.

Thereafter, the participants read the printed instructions placed in each cubicle. To provide assurance that all players were symmetrical the experimenters pointed out that the instructions were also projected onto a wall visible to all. The experiment started after all participants had signaled that they understood the instructions and were ready.

At each of the 80 rounds, each player was randomly assigned to a group of $n=4$ and received a message that she had been allocated $e=100$ tokens to divide between a private account and a group account. After everyone had made their decision, the individual earnings were computed. Each subject received an earnings message in which the information about others' contributions varied by treatment as described in Section 2.4. A new round started after everyone had acknowledged the earnings feedback.

At the end of the session participants returned their written instructions to an experimenter, were asked to not share details of the experiment, and were privately paid their cash earnings.

## 2.6 PAYMENTS AND EARNINGS

The show-up fees and earnings were identical across countries in terms of opportunity cost, by approximating the minimum wage at each location. The experiments were run in late 2014 and early 2015 when the Arizona minimum wage was US$ 8.05/hour. Iceland does not have a minimum hourly wage, but at that time the guaranteed lowest monthly income ("Lágmarkstekjutrygging") was ISK 245,000.00 (Confederation of Icelandic Enterprise, 2014). With a standard 172 monthly work hours this amounts to 1,424 ISK per hour.

The US show-up fee was $5, or 0.62 minimum wage hours. The Icelandic show-up fee was ISK 1000, or 0.7 minimally remunerated hours. In equilibrium where everybody keeps all their tokens in the private account each subject earns 100 tokens per round, 8000 tokens over 80 rounds. The equilibrium earnings over 80 rounds correspond to 1.8 minimum wage hours in both locations.[19]

## 3. RESULTS[20]

We divide our analysis into twelve short results starting with the contribution levels. We conclude with differences and commonalities in the underlying thought processes.

---

[19] In Iceland, the conversion rate was 0.32 ISK per token; ISK0.32 * 8000 = ISK 2560, or 1.8 minimally remunerated hours. The US conversion rate was 0.18 cents per token, US$ 0.0018 * 8000 = US$ 14.40, and 1.8 minimum wage hours.
[20] The raw data in the Supplemental Materials.

**Result 1: Under group ("team") feedback, mean contributions over rounds do not differ between the US and Iceland.**

Fig. 2 depicts the mean contributions per round, country, and treatment. Under group-level feedback (the standard feedback in VCM experiments) there is no overall country difference in the contribution levels,[21] and Hypothesis 1 is not supported. The mean contribution over all 80 rounds is 34 out of 100 tokens both in Iceland and in the US. The mean first round contribution is 47 tokens in Iceland and 43 tokens in the US, both in the range typically reported from Western students (Zelmer, 2003; Ledyard, 1995; Davis & Holt, 1993), and this difference is also not significant.[22]: By round 80 mean contributions have declined to 25 tokens in Iceland and 26 tokens in the US. As mentioned in Section 1.6.3, even different cultural lenses might lead to similar cooperation levels. In Results 5, 11, and 12 we show that different cultural lenses underlie these superficially similar contribution patterns.

***Figure 2**. Mean contribution per round, country, and treatment.*

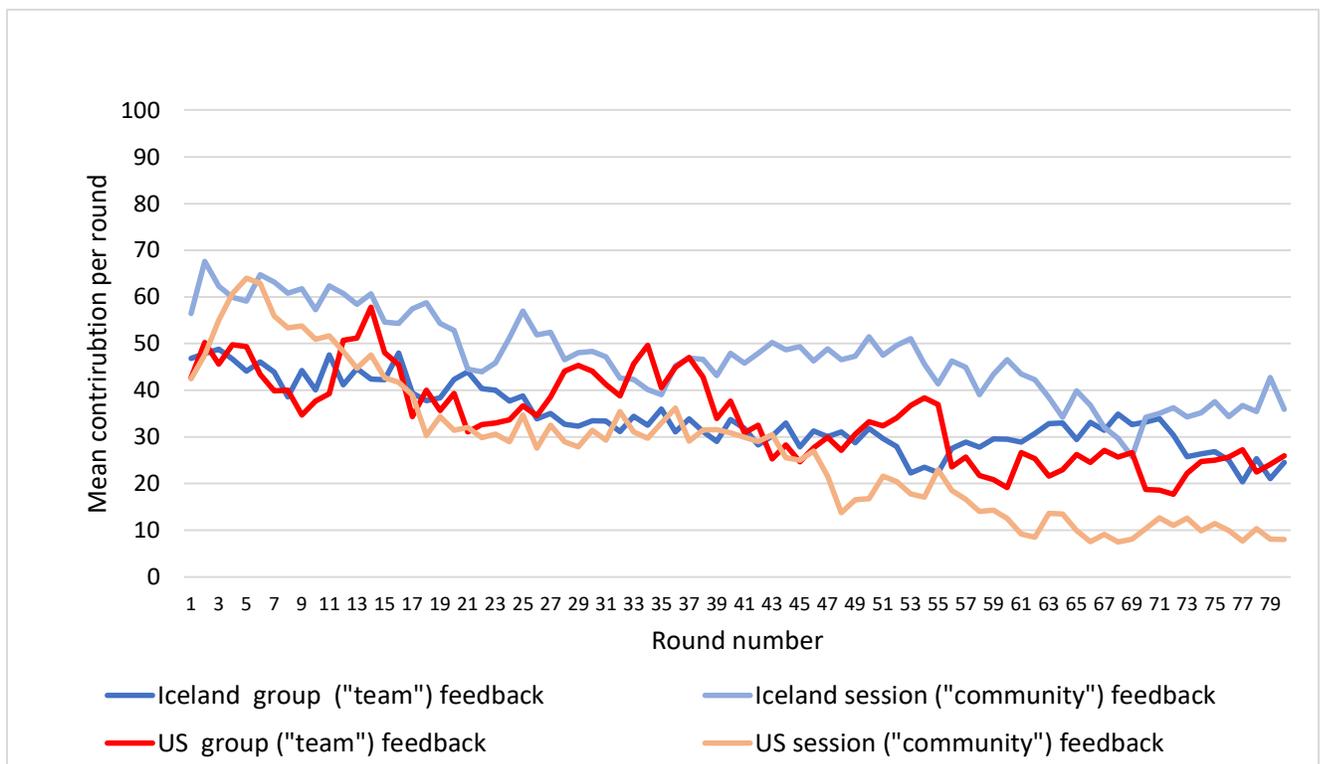

---

[21] Mann-Whitney-Wilcoxon U (henceforth, **MWU**) normal approximation (Siegel & Castellan, 1988), $n=m=36$, $z=0.15$, $p$(2-tailed) = 0.44; the unit of analysis is an individual subject's contributions over all rounds.
[22] $t(70)=0.65$, $p$(2-tailed)= 0.52.

**Result 2: Under session ("community") feedback contributions are higher in Iceland than in the US.**

Informing participants of the contribution of everyone in the session and thus focusing them on a wider community (all session participants), results in a clear country difference (Fig. 2). This supports Hypotheses 1 and 3. It is also suggestive of Hypothesis 2 since we observe culturally different responses to what is effectively a priming message about membership in a wider social unit. Recall that the participants know from the instructions, and thus at the very start of Round 1 before they have seen the first post-round feedback, that everybody's contribution will be viewed by all others in the session, albeit anonymously. Already in Round 1, the Icelandic mean contribution of 56 tokens significantly exceeds the US mean of just 42 tokens.[23] Over all 80 rounds, the mean contribution is 56 tokens in Iceland but only 27 tokens in the US. In Round 80, Icelanders still contribute 36 tokens on average while the corresponding US value is only 8 tokens. The overall difference between contributions in Iceland and the US is highly significant: Mann-Whitney U (henceforth, **MWU**) normal approximation, $m = 36$, $n = 48$, $z=3.84$, $p$(2-tailed) <0.000.[24]

**Result 3: In Iceland, session ("community") feedback raises contributions compared to group ("team") feedback. This effect is present in the earliest rounds and is stable over rounds.**

Further in accordance with Hypothesis 3, Fig. 2 shows that the Icelandic mean contribution per round under community feedback exceeds the Icelandic team feedback mean in nearly every[25] round. Over all rounds, the mean contribution is 56 tokens under community feedback and 47 tokens under group feedback. MWU tests reject the null hypothesis that the contribution levels are the same in the two feedback conditions.[26] Already in the first round there is a significant difference in contributions.[27] The fact that the difference in cooperativeness between feedback treatments emerges in the very first round rather than gradually suggests that not just the session-wide post-round feedback but already the treatment-specific instructions focus Icelandic subjects on their wider social unit, and that this primes some sort of a schema, possibly a morally driven community focus.

---

[23] $t(82) = 2.47$, $p$(1-tailed) <0.01.
[24] Using Jonckheere tests (Siegel & Castellan, 1988) we further tested the hypothesis that within each round, the medians under group feedback and community feedback are the same against the alternative that the medians of the community feedback distributions are higher. We find in favor of the alternative at the 5 percent significance level in 60 out of 80 rounds.
[25] Rounds 68 and 69 excepted.
[26] MWU normal approximation, $m=36$, $n=48$, $z=2.92$, $p$(1-tailed) <0.002, the unit of analysis is the individual contribution over rounds.
[27] Using Jonckheere tests we also examined the hypothesis that within each round, the medians under group feedback and community feedback are the same against the alternative that the medians under community feedback are higher. We find in favor of the alternative at the 5 percent significance level already in the first ten rounds, and overall, in 52 of the 80 rounds.

**Result 4: For US subjects, session-level feedback leads to lower contributions in the later rounds of the experiment.**

For US subjects, the effect of community feedback trends opposite of what is observed in Iceland. The mean contribution over sessions and rounds is lower under community feedback (27 tokens) than under group feedback (34 tokens). Taken over all 80 rounds this difference is not statistically significant.[28] However, a closer look at Fig. 2 reveals a change over time: The contribution levels of group and community feedback are similar until Round 21, and even intersect: In Rounds 3 – 11, 17, and 21 the mean under community feedback exceeds the corresponding group feedback mean. Thereafter however, group feedback means reliably exceed community feedback means, with the sole exception of round 43. If only Rounds 41 to 80 are considered[29] the difference in contribution levels between feedback treatments is statistically significant.[30] The gradual emergence of a difference suggests that it is due to learning and experience. This finding is unexpected but Result 5 helps uncover the reason for this prime-driven effect and provides evidence in support of Hypothesis 2 that the two cultures think differently about cooperation.

***Table 3***. *Correlations between own contribution and the contributions by the other group members in the preceding round, and between the number of free riders encountered in one's group in the preceding round.*

| Location & treatment | Iceland group feedback | US group feedback | Iceland community feedback | US community feedback |
|---|---|---|---|---|
| Number of subjects | 36 | 36 | 48 | 36 |
| Mean individual correlation over all rounds | 0.14 | 0.20 | 0.22 | 0.44 |
| Correlation between $x_{it}$ and the number of free riders encountered in one's group in Round $t$-1 | -0.08 | -0.20 | -0.13 | -0.42 |

**Result 5: US students reciprocate more; This is associated with lower cooperation rates. The effect is most pronounced under community feedback in later rounds.**

Many VCM studies show that a sizeable fraction of subjects are conditional cooperators (e.g., Croson, 2007; Page, Putterman & Unel, 2005; Keser & Van Winden, 2000). If a sufficient proportion of players adopt such strategies it can slow the robustly observed decay in contributions over rounds. Reciprocity does not fully halt the decay because of variability both in the contributions a player observes, and in the degree to which individuals match others' contributions if at all. With sufficiently many free riders present, reciprocity can even accelerate the decline in contributions (Gunnthorsdottir et al., 2007; Page et al.).

---

[28] $m = n = 36$, $z = 1.2$, $p$(2-tailed) = 0.22. (MWU, normal approximation, the unit of analysis is an individual subject's mean contribution over rounds.)

[29] Analyzing contributions in blocks of rounds, as here, assesses learning over time. Behavior typically stabilizes in later rounds (e.g., Duffy & Hopkins, 2005).

[30] MWU normal approximation, $m=n=36$, $z = 2.22$, $p$(2-tailed) = 0.03.

Are the different trends in contributions over rounds in Fig. 2 associated with differences in reciprocity? As an initial step to compare the extent of conditional cooperation between cultures and feedback treatments, we compute each subject's correlation, over all rounds, between his contribution in Round $t$ and the contributions by the three other members of his group in Round $t$-1. We then calculate the mean of individuals' correlations, per treatment and country. These means are displayed in the second row of Table 3:

1) In both feedback treatments, the mean correlations are higher in the US than in Iceland. Under group ("team") feedback the country difference is neither large nor statistically significant.[31] Under community feedback, the difference is both significant and substantial.[32]
2) In both countries, the mean correlations are significantly[33] higher under community feedback than under group feedback, but this difference is much larger for US subjects.
3) The US/community feedback cell, with the fastest decline in contributions over rounds (Fig. 2), has the highest mean correlation among all four experimental cells.

To confirm the robustness of these findings about differences in reciprocity, we next apply an alternative approach to the same question: Following Isaac and Walker's (1988) seminal classification, we define an individual who contributes $< 1/3e$ (here, 33 tokens or less) as a "free rider" in that round. We then compute the correlation between how many free-riders a subject encountered in her group in Round $t$-1 and her contribution in Round $t$. Since this is a robustness check we also compute the correlations differently from before: Instead of calculating the per-cell mean of individual subjects' correlations we compute one correlation per cell, over subjects and rounds. The pattern of differences between the cells in Row 3 of Table 3 resemble those in Row 2. We assess the statistical significance of all pairwise differences with a Fisher $r$ to $z$ transformation (Cohen & Cohen, 1983, Ch. 2). The results echo the prior analysis:

1) In both feedback treatments, the US negative response to encounters with free riders is stronger than in Iceland. This cultural difference, statistically significant in both feedback treatments,[34] is greatest under community feedback.

---

[31] MWU normal approximation, $m=n=36$, $z=1.40$, $p$(2-tailed) = 0.16.
[32] $m=36$, $m=48$, $z = 3.53$, $p$(2-tailed) = 0.00.
[33] For Iceland, $m=36$, $n=48$, $z=1.97$, $p$(2-tailed) = 0.05, for the US, $m=n=36$, $z=3.66$, $p$(2-tailed) = 0.00.
[34] For group feedback, $z = 2.03$, $p$(2-tailed) = 0.02, for community feedback, $z=12.77$, $p$(2-tailed) = 0.00.

2) In both countries, the response to encounters with free riders in one's group is stronger under community feedback than under group feedback. Both pairwise differences are significant.[35]

3) Among the four experimental cells the reciprocal response to encountering free riders in one's group is the strongest among US subjects who received session ("community") feedback.

**Result 6: In both treatments, the structure of the contribution decisions differs between the two cultures.**

Result 5 provides initial support for Hypothesis 2, that the two cultures think differently about cooperation. However, what ultimately motivates Icelandic subjects to cooperate at an equal or even higher level than US subjects even though they are less reciprocal? Could there be different cultural lenses, equally effective, that drive cooperation? To address this, we conduct a detailed analysis of the factors that impact contribution decisions in the two cultures:

Table 4 shows double-censored Tobit regressions of individual group contributions on possible decision factors, by country and feedback treatment. We build on Ashley et al.'s (2010) pioneering econometric analysis of motives to cooperate. We add first ($t=1$) round contributions as an explanatory variable since they are an indicator of stable individual dispositions (Gunnthorsdottir et al. 2007; Ones & Putterman, 2007). Since there is a subject effect and individuals' residuals are correlated over rounds, we cluster the standard errors by participant. This does not affect the size of the coefficients or the model's overall fit but robustly guards against Type I errors (Petersen, 2009).[36] Table 5 displays the differences between the regression coefficients across cultures. Shaded cells contain statistically significant differences where $p < 0.05$. We apply the most conservative approach to the computation of the significance levels of these differences (Clogg, Petkova & Haritou, 1995; Paternoster et al., 1998) where

$$z = \frac{\beta_1 - \beta_2}{\sqrt{SE\beta_1^2 + SE\beta_2^2}} \qquad \text{(Eq. 2)}$$

and since we do not have grounds for directional hypotheses all tests are two-tailed. A mere glance at the shaded cells in Table 5 reveals that the decision to cooperate is structurally different between the two cultures, providing further support for Hypothesis 2. We next explore Tables 4 and 5 in detail, reporting first on cross-cultural commonalities (Result 7-10), and then on cultural differences (Result 11, 12).

---

[35] For Iceland, $z=2.04$, $p$(2-tailed) =0.04; for the US, $z=0.9.23$, $p$(2-tailed) = 0.00.

[36] In a widely cited paper, Petersen (2009) compares different methods to analyze panel data when the residuals are correlated across groups, individuals, or time. We also tried different econometric models and like Petersen, find clustered standard errors to be the most conservative approach. Other common methods produced smaller standard errors (and, spurious significance of coefficients).

*Table 4*. Double censored Tobit regression of group contributions in round t on decision factors. Standard errors clustered by individual, standard errors in parentheses.

|  | **Iceland group feedback** | **US group feedback** | **Iceland community feedback** | **US community feedback** |
|---|---|---|---|---|
| Intercept | -.11.56** (4.38) | -32.67** (7.49) | -5.78 (6.77) | -22.85** (5.35) |
| Own contribution in Round $t=1$ | 0.28** (0.06) | 0.34** (0.09) | 0.16 (0.12) | 0.19** (0.06) |
| Own contribution in Round $t-1$ | 0.70** (0.07) | 1.11** (0.11) | 0.67** (0.05) | 1.07** (0.10) |
| Own contribution in Round $t-2$ | 0.25** (0.04) | 0.26** (0.08) | 0.39** (0.05) | 0.43** (0.06) |
| Over-contribution in Round $t-1$ relative to the group mean [@] | -0.31** (0.10) | -0.42** (0.07) | -0.24** (0.08) | -0.61** (0.12) |
| Under-contribution in Round $t-1$ relative to the group mean [@] | +0.10 (0.06) | +0.23** (0.09) | -0.01 (0.07) | +0.24** (0.08) |
| Count of zero contributors in the session in Round $t-1$[@] |  |  | -1.04** (0.52) | -1.27** (0.49) |
| Count of full contributors in the session in Round $t-1$[@] |  |  | -0.85 (0.63) | -1.19 (0.65) |
| $N$ | 2808 | 2808 | 3744 | 2808 |
| McFadden Pseudo $R^2$ | 0.07 | 0.10 | 0.08 | 0.13 |
| LLF | -11,043.84 | -8,574.38 | -13,994.88 | -7,895.32 |
| Correlation observed - predicted | 0.65 | 0.69 | 0.72 | 0.79 |
| % censored at 0 | 15% | 34% | 16% | 40% |
| % censored at 100 | 4% | 12% | 14% | 8% |

[@]excludes $i$
** $p < 0.01$, * $p < 0.05$ (both 2-tailed).

**Result 7: In both countries and in both treatments, subjects are in part guided by a stable personal heuristic.**

Subjects bring to the laboratory their habitual behaviors and frames (Henrich et al., 2004, p. 46; Hoffman, McCabe & Smith, 1998). In a VCM, this includes their cooperative style. The latter is known to be heterogeneous even within a culture, and includes unconditional cooperation, conditional cooperation, free-riding, or hybrid strategies (Cherry et al., 2008; Croson, 2007; Kurzban & Houser, 2005; Isaac & Walker, 1988). We capture players' cooperative style by their contribution in Round $t=1$, and their behavioral stability over rounds by their contributions in Rounds $t-1$ and $t-2$.

***Table 5***. *Coefficient differences between the two cultures and their statistical significance computed according to Eq. (2). The US coefficients are subtracted from the coefficients for Iceland. Cells with significant differences (p≤0.05) are shaded.*

|  | **Group feedback** |  | **Community feedback** |  |
| --- | --- | --- | --- | --- |
|  | Iceland minus US | z | Iceland minus US | z |
| Intercept | 21.11 | 2.43* | 17.07 | 1.98* |
| Own contribution in Round $t$=1 | -0.06 | 0.57 | -0.03 | 0.22 |
| Own contribution in Round $t$-1 | -0.41 | 3.17** | -0.40 | 3.58** |
| Own contribution in Round $t$-2 | -0.01 | 0.11 | -0.04 | 0.52 |
| Over-contribution in Round $t$-1 relative to the group mean | 0.11# | 0.88 | 0.37# | 2.60** |
| Under-contribution in Round $t$-1 relative to the group mean | -0.13## | 1.20 | -0.25## | 2.40* |
| Count of zero contributors in the session in Round $t$-1 |  |  | 0.23 | 0.32 |
| Count of full contributors in the session in Round $t$-1 |  |  | 0.33 | 1.20 |

\* $p < 0.05$, \*\* $p < 0.01$, all two-tailed
\# After over-contributing at $t$-1 Icelandic subjects lower their contribution by less.
\#\# After under-contributing at $t$-1 Icelandic subjects raise their round $t$ contribution by less.

Table 4 shows that the first round ($t$=1) contribution is an indicator of one's contribution level throughout the experiment, in both cultures and both treatments. Similar results have been reported elsewhere (Gunnthorsdottir et al., 2007; Ones & Putterman, 2007). There is no country difference in how the $t$=1 contribution predicts subsequent behavior (Table 5). Furthermore, the coefficients of contributions at Rounds $t$-1 and $t$-2 are significant in all four columns of Table 4. Table 5 shows that the coefficients of $t$-1 contributions are lower in Iceland than in the US in both treatments, but there is no country difference in the impact of $t$-2 contributions. The finding that US subjects' $t$-1 contributions have a higher impact on their subsequent decision in Round $t$ might reflect individual stability and independence from the group. It could be linked to the fact that the US have the most individualist culture in the world (Hofstede, Hofstede & Minkov, 2010, p. 97).[37]

---

[37] This cross-cultural coefficient difference, which we did not expect, requires investigation beyond the scope of this paper.

**Result 8: In both locations and in both treatments, subjects consider the mean of other group members' contributions in the preceding round.**

Each 80-round experiment lasted just under 90 minutes, about a minute per round. Without formal time constraints subjects in both countries settled on this relatively fast pace. In addition to their individual type-based heuristics (Result 7) subjects must rely on information that they can process in such limited time. Recall that under group feedback only the individual contributions within one's group were displayed. Under community feedback where all session participants' contributions were displayed, one's own group visually stood out from among the other groups. In both feedback treatments, subjects could thus quickly process the minimum, maximum and possibly, median contribution in their own group of four. In addition to the display of individual contributions, in both treatments the sum of the contributions by the other three group members was shown in a separate message box, so that it was possible to get a sense of their mean. There was also a reminder of one's own contribution, allowing a comparison.

Like Ashley, Ball & Eckel (2010), we separate individual deviations from the mean contribution by the other three group members at Round $t$-1 into two independent variables (Table 4): "Over-rcontribution in Round $t$-1 relative to the group mean" is the amount by which an individual's contribution exceeded the mean of the other three group members, zero otherwise. "Under-contribution in Round $t$-1 relative to the group mean" similarly captures under-contribution. Table 4 shows that participants in both countries and in both treatments are guided by the mean of other group members' contributions in the preceding round. The coefficients for over-contribution and under-contribution in Round $t$-1 are highly significant. The sole exception is under-contribution relative to others in the Icelandic community feedback treatment.[38]

**Result 9: In both cultures, players whose contribution in Round $t$-1 had been below the mean of what the other group members had contributed made a smaller reciprocal adjustment than players whose contribution had been above the mean. This is consistent with Inequity Aversion.**

Inequity Aversion (Fehr & Schmidt, 1999; Ashley et al., 2010) appears cross-cultural in our data. Table 4 shows that consistent with Fehr & Schmidt and many other studies since, players in both countries reciprocate asymmetrically: Comparing the coefficients for "Over-contribution" and "Under-contribution", the downward adjustment after over-contributing at Round $t$-1 relative to the other group members is two to three times as large as the upward adjustment after under-contributing. In Iceland under community feedback there is no upward adjustment at all. This difference in the size of upward and downward adjustments can be accounted for by the alignment, or lack thereof, between two motives: Under-contribution is congruent with self-interest but runs counter to reciprocity/fairness; over-contribution runs counter to both motives (see Eq. 1). It will be interesting to eventually see whether Inequity Aversion is a human universal.

---

[38] The reader might wonder why subjects did not consider a simpler value directly or near-directly displayed on the screen such as the group maximum, minimum, or median. Additional analyses reject this. Participants in both cultures focused on the mean contribution of the other group members.

**Result 10: In both locations, subjects consider the frequency of zero contributors in the session if this information available.**

Under community feedback, subjects in *neither* country considered the central tendency of everybody's contribution in the preceding round. They instead relied on a simpler metric, the count of zero contributors in Round $t$-1. The associated coefficients (Table 4) are highly significant in both cultures, and not culturally different (Table 5). The more zero contributors a subject observed the more she reduced her contribution in the round that followed. Subjects might have gone with an absolute count rather than the value of the minima or maxima because their variability was low,[39] so that their value would have provided little reason to adjust one's contribution in the round that followed.

**Result 11: Under community feedback, US subjects make stronger reciprocal adjustments than Icelanders do.**

Table 5 shows that if informed of the actions of everybody in the session, US subjects make stronger adjustments following their under-contribution or over-contribution in the preceding round than Icelanders do. The coefficient difference is not significant under group feedback, but it is significant under community feedback, as already indicated in Results 4 and 5 above. The coefficient difference is further support for Hypothesis 2 that Americans think differently about cooperation than Icelanders do.

**Result 12: Icelandic subjects have a stronger disposition to contribute unconditionally.**

As further confirmation of Hypothesis 2, Icelanders have a stronger tendency to cooperate unconditionally. This is reflected in the intercept difference (Table 4). The intercepts capture cooperative tendencies absent of experience, individual cooperative style (captured by contribution in $t$=1) or knowledge of others' actions. The country difference between the intercepts is statistically significant and substantial in both treatments (Tables 4, 5). Pairwise comparisons of the intercepts between feedback treatments but within cultures reveal no significant within-culture differences ($z=0.72$ for Iceland, $z=1.07$ for the US). The intercepts are thus culture specific and not related to the nature of the feedback. Recall also that Icelanders contribute more under community feedback than under group feedback starting in Round 1 (Result 3), which suggests a principle-based community focus.

---

[39] In Iceland, in 89% of the rounds, the session maximum was 80 tokens or higher. In 76% of the rounds the maximum was 100 tokens. In the US, 60% of the maxima were ≥ 80 tokens, with 45% at 100 tokens. In Iceland, the round minimum under community feedback was zero in 78% of the rounds, and ≤10 tokens in 92% of the rounds. In the US, 94% of the round minima were zero.

## 4. DISCUSSION

*Cross-cultural similarities.* We had not hypothesized about cross-cultural similarities but found some, including individual consistency over rounds (Result 7), being guided by other group members' contributions (Result 8) and the count of non-contributors in the wider social unit (Result 10), and maybe most interestingly, Inequity Aversion (Result 9). It remains to be seen whether these commonalities are ultimately culture-specific and shared between the US and Iceland, or whether they are human universals.

*Hypothesis 1.* Our conjecture that contribution levels differ between the two nations because they belong to different clusters on Inglehart's world map of cultures, is confirmed under community feedback (Result 2) but not under group feedback (Result 1).

*Hypothesis 2.* The two subject pools are superficially similar: Students from both nations qualify as WEIRD (Western, Educated, Industrialized, Rich, Democratic; Henrich et al., 2010) and the countries' GDP/capita are very close.[40] Under group feedback there is not even a difference in contribution levels. Yet, Hypothesis 2 is confirmed: Whether cooperation levels differ or not, the motives to cooperate differ between the two cultures (Result 6). US students are intense reciprocators who tend to match the contributions of those they directly cooperate with (Results 5, 11).[41] When US subjects are informed of the contributions not only of their team but of the mini community of session participants, their reciprocal focus is sharpened, which accelerates the decline in contributions over time (Result 4). Icelanders in contrast are more inclined to cooperate unconditionally. They act as if on principle and are less sensitive to the contributions of others in their group, especially when given a community focus (Results 3, 12).

*Hypothesis 3.* The greater community focus of Icelanders reflected in their higher score on Knack & Keefer's (1997) scale of civic values gave rise to Hypothesis 3, that they cooperate more if primed about membership in wider social unit than just their team. This is confirmed (Results 2, 3). The cultural divergence in the impact of the community priming message provides additional support for Hypothesis 2 that the two cultures apply different cognitive lenses to cooperation.

---

[40] PPP adjusted, close both in international rank and in absolute terms (OECD, 2020; World Bank, 2019)
[41] Cherry et al. (2008) use a different method but report a similar result: Their US subjects were more frequently conditional cooperators than their Japanese or Austrian subjects.

## 5. CONCLUSION

The finding that the motives to voluntarily cooperate differ between subjects in Iceland and the US suggests practical applications on how to enhance cooperation in each culture. Icelanders might respond best to moral suasion aimed at a sense of community and social obligation while Americans might be more responsive to appeals reciprocity and fairness. This is indicated by differences we uncovered in the underlying rationales for cooperation, as well as by the finding that a wider community focus increases cooperation in Iceland but decreases it in the US.

Culture reflects a society's history and is often persistent (see e.g., Gershman, 2017 for an overview). The differences in motives to cooperate that we found appear to fit with the two nations' longstanding demographic, social and geographic conditions: Iceland's small homogenous population has long been stationary on an island. Even people who do not directly know each other often at least know *about* each other. Reputation or "image scoring" (Wedekind & Milinski, 2000; Nowak & Sigmund, 1998) matters for success in a small, stationary society. A heuristic to cooperate unconditionally is functional both for the individual and the collective. In contrast, in a large, mobile, heterogeneous population such as the US, unconditional cooperators benefit less from being known as such. It would make them vulnerable to rational free riders who are unconcerned about the long-term reputational impact of their selfish acts because they can easily move away. In a large mobile society keeping a close eye on others' cooperative input and directly reciprocating others' action is both individually and collectively functional.

### 5.1 FURTHER RESEARCH

The West and East Asia are culturally even more different than Iceland and the US. Their different cognitive and perceptual habits are well demonstrated (see Nisbett & Masuda, 2003, for an overview). One region is highly individualist the other collectivist (Hofstede, Hofstede & Minkov, 2010). However, multiple experimental studies (Section 1.6.3) report similar cooperation levels in these two cultural areas. It would therefore be interesting to compare which factors impact VCM contributions in the West and East Asia. Since many VCM experiments have been run with reciprocity focused US subjects it remains to be seen whether reciprocity is as common as hitherto assumed or whether there exist other equally effective cultural norms that encourage voluntary cooperation.

Since we also find cross-cultural commonalities in how cooperation decisions are processed across the two cultures (Inequity Aversion, personal consistency) it would be worthwhile to explore whether and to which extent these patterns are human universals or ultimately a feature of culture.

# REFERENCES

Ahn, T. K. & Elinor Ostrom (2008). The meaning of social capital and its link to collective action. In D. Castiglione, Jan W Van Deth, & Guglielmo Wolleb (eds.), *The Handbook of Social Capital.* Oxford, UK: Oxford University Press. Ch. 3, pp. 70-100.

Alesina, Alberto, & Nicola Fuchs-Schündeln (2007). Goodbye Lenin (or not?): The effect of communism on people. *American Economic Review 97*(4), 1507-1528.

Algan, Yann, & Pierre Cahuc (2010). Inherited trust and growth. *American Economic Review 100*(5), 2060-2092.

Andreoni, James (1995). Warm-glow versus cold-prickle: the effects of positive and negative framing on cooperation in experiments. *Quarterly Journal of Economics 110*, 1–21.

Andreoni, James, & Rachel Croson (2008). Partners versus strangers: Random rematching in public goods experiments. In Vernon Smith & Charles Plott (eds.), *Handbook of Experimental Economics Results Vol. 1.* Ch. 82, p. 776-782. Amsterdam: North-Holland.

Ashley, Richard, Sheryl Ball, & Catherine Eckel, (2010). Motives for giving: A reanalysis of two classic public goods experiments. *Southern Economic Journal 77*(1), 15-26.

Banfield, Edward C. (1958). *The Moral Basis of a Backward Society*. NY: The Free Press.

Bigoni, Maria, Stefania Bortolotti, Marco Casari, Diego Gambetta, & Francesca Pancotto (2016). Amoral familism, social capital, or trust? The behavioural foundations of the Italian North–South divide. *Economic Journal 126*(594), 1318-1341.

Blackwell, Calvin, & Michael McKee (2010). Is there a bias toward contributing to local public goods? Cultural effects. *Forum for Social Economics*, *39*(3), 243-257.

Boyd, Robert, & Peter J. Richerson (2002). Group beneficial norms spread rapidly in a structured population. *Journal of Theoretical Biology* 215, 287–296.

Boyd, Robert, & Peter J. Richerson (1985). *Culture and the Evolutionary Process*. Chicago: Univ. of Chicago Press.

Brandts Jordi, Tatsuyoshi Saijo, & Arthur Schram (2004). How universal is behavior? A Four Country Comparison of Spite and Cooperation in Public Goods Games. *Public Choice 119*, 381- 424.

Camerer, Colin (2003). *Behavioral Game Theory: Experiments in Strategic Interaction*. Princeton, NJ: Princeton University Press.

Castro, Massimo F. (2008), Where are you from? Cultural differences in public good experiments. *Journal of Socio-Economics 37*(6), 2319-2329.

Cavalli-Sforza, Luigi Luca, & Marcus William Feldman (1981). *Cultural Transmission and Evolution: A Quantitative Approach*. Princeton, NJ: Princeton University Press.

Cherry, Todd, Martin Kocher, Stephan Kroll, Robert Netzer, & Matthias Sutter (2008). Conditional cooperation on three continents. *Economics Letters 101*(3), 175-178.

Chen, Kang, & Fang-Fang Tang (2009). Cultural differences between Tibetans and Ethnic Han Chinese in ultimatum game bargaining experiments. *European Journal of Political Economy 25*, 78–84.

Clogg, Clifford, Eva Petkova, & Adamantios Haritou (1995). Statistical methods for comparing regression coefficients between models. *American Journal of Sociology 100*(5), 1261-1293.

Cohen, Jacob, & Patricia Cohen (1983). *Applied Multiple Regression/Correlation Analysis for the Behavioral Sciences* 2nd Ed. London: Erlbaum.

Coleman, James S. (1988). Social capital in the creation of human capital. *American Journal of Sociology* 94S, S95-S120.

Confederation of Icelandic Enterprise 2014 [cited 14 March 2015]. In Kaupgjaldskrá SA [Internet]. Reykavik: Confederation of Icelandic Enterprise. Available from: http://sa.vinnumarkadur.is/kaupgjaldsskra-sa.

Cox, Caleb A., & Brock Stoddard (2015). Framing and feedback in social dilemmas with partners and strangers. *Games 6*, 394–412.

Cronk, Lee (1999). *That Complex Whole: Culture and the Evolution of Human Behavior.* Boulder, CO: Westview.

Croson, Rachel (2007). Theories of commitment, altruism and reciprocity: Evidence from linear public goods games. *Economic Inquiry 45*, 199-216.

Davis, Douglas, & Charles Holt (1993), *Experimental Economics*. Princeton, NJ: Princeton University Press.

Douglas, Mary, & Aaron Wildavsky (1983). *Risk and Culture: An Essay on the Selection of Technological and Environmental Dangers*. Berkeley: University of California Press.

Duffy, John & Ed. Hopkins (2005). Learning, Information and Sorting in Market Entry Games: Theory and Evidence. *Games and Economic Behavior 51*, 31–62.

Eckel, Catherine, Haley Harwell, & Jose G. Castillo (2015). Four classic public goods experiments: A replication study. *Replication in Experimental Economics* (*Research in Experimental Economics 18)*, 13-40.

Ehmke, Mariah Tanner, Jayson Lusk, & Wallace Tyner (2010). Multidimensional tests for differences in economic behavior across cultures. *Journal of Socio-Economics 39*(1), 37-45.

Ensminger, Jean, & Joseph Henrich (Eds.) (2014). *Experimenting with Social Norms: Fairness and Punishment in Cross-Cultural Perspective.* NY: Russell-Sage.

Fernández, Raquel (2011). Does culture matter? In, Jess Benhabib, Alberto Bisin, & Matthew O. Jackson (eds.), *Handbook of Social Economics, Vol. 1A,* Amsterdam: North-Holland, Ch. 11.

30 of 36Page **30** of **36**

Fernández, Raquel (2008). Culture and economics. In: Palgrave Macmillan (eds.) *The New Palgrave Dictionary of Economics*, London: Palgrave Macmillan.

Fukuyama, Francis (2001). Social capital, civil society and development. *Third World Quarterly 22*(1), 7– 20.

Gächter, Simon, & Benedikt Herrmann (2009). Reciprocity, culture and human cooperation: Previous insights and a new cross-cultural experiment. *Philosophical Transactions of the Royal Society B 364*(1518), 791–806.

Gächter, Simon, Benedikt Herrmann, & Christian Thöni (2010). Culture and cooperation. *Philosophical Transactions of the Royal Society B 365*, 2651–2661.

Garreau, Joel (1981). *The Nine Nations of North America*. Boston, MA: Houghton Mifflin.

Gershman, Boris (2017), Long-run development and the new cultural economics. In Matteo Cervelatti and Uwe Sunde (eds.), *Demographic Change and Long-Run Development*. Cambridge, MA: MIT Press. Ch. 9, pp. 221–261.

Gunnthorsdottir, Anna, Dan Houser, & Kevin McCabe, K. (2007), Disposition, history, and contributions in public goods experiments. *Journal of Economic Behavior and Organization 62*(2), 304-315.

Guiso, Luigi., Paolo Sapienza, & Luigi Zingales (2006). Does culture affect economic outcomes? *Journal of Economic Perspectives 20*(2), 23–48.

Guiso, Luigi., Paolo Sapienza, & Luigi Zingales (2004). "The role of social capital in financial development. *American Economic Review 94*(3), 526-56.

Gurven, Michael (2004). Does market exposure affect economic game behavior? In Joseph Henrich et al. (Eds.), *Foundations of Human Sociality* (pp. 194–231). NY: Oxford University Press.

Hardin, Garrett (1968). The tragedy of the commons. *Science 162*, 1243–1248.

Henrich, Joseph, Robert Boyd, Samuel Bowles, Colin Camerer, Ernst Fehr, Herbert Gintis, Richard McElreath, Michael Alvard, Michael, Abigail Barr, Jean Ensminger, Natalie Smith Henrich, Kim Hill, Francisco Gil-White, Michael Gurven, Frank W. Marlowe, John Q. Patton, & David Tracer (2005). Economic man' in cross-cultural perspective: behavioral experiments in 15 small-scale societies *Behavioral and Brain Sciences 28*, 795-855.

Henrich, Joseph, Robert Boyd, Samuel Bowles, Colin Camerer, Ernst Fehr, & Herbert Gintis, H., (2004) *Foundations of human sociality: Economic experiments and ethnographic evidence from fifteen small-scale societies*. Oxford University Press.

Henrich, Joseph, Robert Boyd, Samuel Bowles, Colin Camerer, Ernst Fehr, Herbert Gintis, & Richard McElreath (2001). In search of homo economicus: Behavioral experiments in 15 small-scale societies. *American Economic Review 91*(2), 73–78.

Henrich Joseph, & Jean Ensminger (2014). Theoretical foundations: the co-evolution of social norms, intrinsic motivations, markets, and the institutions of complex societies. In Jean Ensminger &

Joseph Henrich (eds.), *Experimenting with Social Norms: Fairness and Punishment in Cross-cultural Perspective.* Ch. 2. NY: Russell-Sage.

Henrich, Joseph, Steven J. Heine, & Ara Norenzayan (2010). The weirdest people in the world? *Behavioral and Brain Sciences 33*(2-3), 61-83.

Herrmann, Benedikt, Christian Thöni, & Simon Gächter (2008). Antisocial punishment across societies. *Science* 319, 1362-1367.

Hoffman, E., K. McCabe & V. Smith (1998). Behavioral foundations of reciprocity: Experimental economics and evolutionary psychology. *Economic Inquiry 36*(3), 335-352.

Hofstede, Geert (2001). *Culture's Consequences: Comparing Values, Behaviors, Institutions, and Organizations Across Nations.* Thousand Oaks, CA: Sage.

Hofstede, Geert, Gert Hofstede, & Michael Minkov (2010). *Cultures and Organizations: Software of the Mind*, 3rd Edition. NY: McGraw-Hill.

House, Robert J., Paul Hanges, Mansour Javidan, Peter Dorfman, & Vipin Gupta (2004). *Culture, Leadership, and Organizations: The GLOBE Study of 62 Societies*. Thousand Oaks, CA: Sage.

Inglehart, Ronald (1997). *Modernization and Postmodernization: Cultural, Economic and Political Change in 43 Societies*. Princeton, NJ: Princeton University Press.

Inglehart, Ronald, & Wayne E. Baker (2000). Modernization, cultural change, and the persistence of traditional values. *American Sociological Review 65,* 19–51.

Inglehart, Ronald, & Christian Welzel (2005). *Modernization, Cultural Change, and Democracy: The Human Development Sequence*. Cambridge, UK: Cambridge University Press.

World Bank (2019). *Gross national income per capita*. [cited 3 February 2020]. Washington, DC: World Bank. Available from: https://databank.worldbank.org/data/download/GNIPC.pdf

Isaac, R. Marc, Kenneth McCue, & Charles Plott (1985), Public goods provision in an experimental environment. *Journal of Public Economics* 26, 51-74.

Isaac, R. Marc, & James Walker (1988). Group size effects in public goods provision: the voluntary contributions mechanism. *Quarterly Journal of Economics 53*, 179–200.

Isaac, R. Mark, James M. Walker, & Susan H. Thomas (1984). Divergent evidence on free-riding: An experimental examination of possible explanations. *Public Choice 43,* 114-149.

Keser, Claudia, & Frans Van Winden (2000). Conditional cooperation and the voluntary contribution to public goods. *Scandinavian Journal of Economics 102*, 23-39.

Knack, Stephen, & Philip Keefer (1997). Does social capital have an economic payoff: A cross-country investigation. *Quarterly Journal of Economics 112*, 1251 -1288.

Kocher, Martin & Martinsson, Peter & Visser, Martine (2006). Does stake size matter for cooperation and punishment? *Economics Letters* 99. 508-511.

Kurzban, Robert, & Dan Houser (2005). Experiments investigating cooperative types in humans: A complement to evolutionary theory and simulations. *Proceedings of the National Academy of Sciences 102*(5), 1803-1807.

La Porta Rafael, Florencio Lopez-de-Silanes, Andrei Shleifer & Robert W. Vishny (1997). Trust in large organizations. *American Economic Review 87*(2), 333-338.

Ledyard, John O. (1995), Public goods: A survey of experimental research. In *Handbook of Experimental Economics*, John Kagel, John & Alvin Roth (eds.), 111-194. Princeton, NJ: Princeton University Press.

Morris, Michael W., Richard E. Nisbett, & Kaiping Peng (1995) Causal attribution across domains and cultures. In Dan Sperber, David Premack, & Ann J. Premack (eds.), *Symposia of the Fyssen Foundation. Causal Cognition: A Multidisciplinary Debate.* NY: Clarendon. pp. 577-614.

Nisbett, Richard & Takahiko Masuda (2003). Culture and point of view. *Proceedings of the National Academy of Sciences 100*(19), 11163-11170.

North, Douglass (1990). *Institutions, Institutional Change and Economic Performance*. Cambridge, UK: Cambridge University Press.

Nowak, Martin. A., & Karl Sigmund (1998). The dynamics of indirect reciprocity. Journal of *Theoretical Biology 194*(4), 561−574.

Ockenfels, Axel, & Joachim Weimann (1999). Types and patterns: An experimental East-West comparison of cooperation and solidarity. *Journal of Public Economics* 71, 275-287.

OECD (2020). Gross domestic product (GDP) (indicator). doi: 10.1787/dc2f7aec-en. Accessed on 22 September 2020.

Olson, Mancur (1965). *The Logic of Collective Action: Public Goods and the Theory of Groups*, Cambridge, MA: Harvard University Press.

Olson, Mancur, & Richard Zeckhauser (1966). An economic theory of alliances. *Review of Economics and Statistics 48*(3), 266-279.

Ones, Umut & Louis Putterman (2007). The ecology of collective action: A public goods and sanctions experiment with controlled group formation, *Journal of Economic Behavior & Organization 62*(4), 495 – 521.

Ostrom, Elinor (1990). *Governing the Commons: The Evolution of Institutions for Collective Action.* Cambridge, UK: Cambridge University Press.

Page, Talbot, Luis Putterman, & Bulent Unel (2005). Voluntary Association in Public Goods Experiments: Reciprocity, Mimicry, and Efficiency. *Economic Journal 115*, 1032-1053.

Pang, Chao, & Samuel Bowles (2006). *Can laboratory experiments explain the cultural difference?* Working paper, Peking University & Santa Fe Institute.

Paternoster, Raymond, Robert Brame, Paul Mazerolle, & Alex Piquero (1998). Using the correct statistical test for equality of regression coefficients. *Criminology 36*(4), 859-866.

Petersen, Mitchell (2009). Estimating Standard Errors in Finance Panel Data Sets: Comparing Approaches *Review of Financial Studies 22*, 435-480.

Pigou, Arthur Cecil (1932/2017). *The Economics of Welfare*. NY: Routledge.

Putnam, R. D. Robert, Robert Leonardi, & Raffaella Nanetti (1993). *Making Democracy Work. Civic Traditions in Modern Italy*. Princeton NJ: Princeton University Press.

Samuelson, Paul (1954). The pure theory of public expenditure. *Review of Economics and Statistics 36*(4)*,* 387-389.

Sandler, Todd, & Keith Hartley (2001). Economics of alliances: The lessons for collective action. *Journal of Economic Literature 39*, 869–896.

Schwartz, Shalom H. (1992). Universals in the content and structure of values: Theoretical advances and empirical tests in 20 countries. *Advances in Experimental Social Psychology 25*, 1-65.

Selten, Reinhard (1967). Die Strategiemethode zur Erforschung des eingeschränkt rationalen Verhaltens im Rahmen eines Oligopolexperiments. In H. Sauermann (ed.), *Beiträge zur experimentellen Wirtschaftsforschung*, pp. 136-168. Tübingen: Mohr.

Siegel, Sidney, & N. John Castellan (1988). *Nonparametric Statistics for the Behavioral Sciences*. NY: McGraw-Hill.

Soltis, Joseph, Robert Boyd, & Peter Richerson, P.J. (1995). Can group-functional behaviors evolve by cultural group selection? An empirical test. *Current Anthropology 36*, 473-83.

Spolaore, Enrico (2014). *Culture and Economic Growth,* Cheltenham: Edward Elgar Publishing.

Social Science Research Institute (2012). *Lífsgildi Íslendinga 2009-2010: Results from the European Values Study*. Reykjavik: University of Iceland.

Storr, Virgil, & Arielle John (2019). Why use qualitative methods to study culture in economic life? In *Experimental Economics and Culture* (*Research in Experimental Economics 19)*, Bingley, UK: Emerald. pp.25-52.

Tabellini, Guido (2010). Culture and institutions: economic development in the regions of Europe. J*ournal of the European Economic Association 8*(4), 677–716.

Thöni, Christian (2019). Cross-cultural behavioral experiments: Potential and challenges. In Arthur Schram & Aljaž Ule (eds.), *Handbook of Research Methods and Applications in Experimental Economics*. Cheltenham, UK: Elgar. Ch. 18, pp.349-367.

Torgler, Benno (2004). Cross-culture comparison of tax morale and tax compliance: Evidence from Costa Rica and Switzerland. *International Journal of Comparative Sociology 45*(1), 17-43.

Voigtländer, Nico, & Hans-Joachim Voth (2012). Persecution perpetuated: The medieval origins of anti-semitic violence in Nazi Germany. *Quarterly Journal of Economics 127*(3), 1339-1392.

Weber, Max (1905/2011). *The Protestant Ethic and the Spirit of Capitalism.* NY: Oxford University Press.

Wedekind, Claus & Manfred Milinski (2000). Cooperation through image scoring in humans. *Science 288,* 850-852.

Weimann, Joachim (1994). Individual behavior in a free riding experiment. *Journal of Public Economics 54*, 185-200.

Welzel, Christian (2013). *Freedom Rising: Human Empowerment and the Quest for Emancipation*. NY: Cambridge University Press.

World Values Survey Association (2020a). *Questionnaire and research topics* [cited 3 March 2021]. In World Values Survey [Internet]. Vienna, Austria: World Values Survey Association. Available from: https://www.worldvaluessurvey.org/WVSContents.jsp

World Values Survey Association (2020b). *Inglehart-Welzel cultural map* [cited 1 October 2020]. In: WVS findings and insights [Internet]. Vienna, Austria. Available from: https://www.worldvaluessurvey.org/WVSContents.jsp

Zelmer, Jennifer (2003). Linear public goods experiments: A meta-analysis. *Experimental Economics 6*(3), 299-310.

# APPENDIX A - INSTRUCTIONS

This is an experiment in decision-making. You have already earned $5. for showing up at the appointed time. If you follow the instructions closely and make decisions carefully, you will make a substantial amount of money in addition to your show-up fee.

**Number of periods**

There will be 80 decision-making periods.

**Participants and their endowments**

There are 12 participants in total. In each period, each individual receives an endowment of 100 experimental tokens.

**The decision task**

In each period, you need to decide how to divide your tokens between two accounts: a **private** account and a **group** account. The group account is joint among all members of the group that you are assigned to in that period.

**How earnings from your two different accounts are calculated in each period:**

- Each token you put in the **private account** stays there for you to keep.

- All tokens that group members invest in the **group account** are added together to form the so-called "group investment". The group investment gets doubled before it is equally divided among all group members. Your group has four members (including yourself).

**A numerical example of the earnings calculation in any given period**:

Assume that in a given period, you decide to put 50 tokens into your private account and 50 tokens into the group account. The other three members of your group together contribute an additional 300 tokens to the group account. This makes the total group investment 350 tokens, which gets doubled to 700 tokens (350 * 2 = 700). The 700 tokens are then split equally among all four group members. Therefore, each group members earns 175 tokens from the group investment (700/4=175). In addition to the earnings from the group account, each group member earns 1 token for every token invested in his/her private account. Since you put 50 tokens into your private account, your total profit in this period is 175 + 50 = 225 tokens.

# HOW EACH DECISION-MAKING PERIOD UNFOLDS AND HOW YOU ARE ASSIGNED TO A NEW GROUP IN EACH OF THE PERIODS

### First, you make your investment decision

Decide on the number of tokens to place in the private and in the group account, respectively. To make a private account investment, use the mouse to move your cursor to the box labeled "Private Account". Click on the box and enter the number of tokens you wish to allocate to this account. Do likewise for the box labeled "Group Account". Entries in the two boxes must sum up to your endowment. To submit your investment, click on the "Submit" button. Then wait until everyone else has submitted his/her investment decision.

### Second, you are assigned to the group that you will be a member of in this period

Once every participant has submitted his or her investment decision you will be assigned to a group with 4 members (including yourself).

### The group assignment proceeds in the following manner:

The computer assigns you to a group of four at random. Random group assignment means that the computer randomly decides to which group you will be assigned to in each period. You cannot influence which group you belong to and your group membership is not predictable.

Recall that group membership is determined anew in each period based on a random procedure. Group membership does not carry over between periods!

### After the group assignment, your earnings for the round are computed

See the numerical example above for details of how earnings are computed after you have been randomly assigned to a group.

### End-of period message

At the end of each period you will receive a message with your total experimental earnings for the period (total earnings = the earnings from the group account and your private account added together). This information appears in your Record Sheet at the bottom of the screen.
**[Condition "session feedback"]:** The Record Sheet will also show the group account contributions of *all* participants in a given round and how the computer has randomly assigned them to groups. Your own contribution will be highlighted. **[Condition "group feedback"]:** The Record Sheet will also show the group account contributions by all four members of your group. Your own contribution will be highlighted.

A new period begins after everyone has acknowledged his or her earnings message.

**At the end of the experiment your total token earnings will be converted into US Dollars at a rate of 0.18 cents per token.**